\documentstyle[psfig,float]{aipproc}

\begin{document}
\hyphenation{calculation bi-layer unique para-meters}
\title{Confinement and Quantization Effects\\in Mesoscopic Superconducting
Structures}

\author{V. V. Moshchalkov, V. Bruyndoncx, E. Rosseel,
L. Van Look, M. Baert, M. J. Van Bael,
T. Puig\thanks{Present address: Institut de Cienca de Materials de Barcelona,
08193 Bellaterra, Spain}, C. Strunk\thanks{Currently at: University of Basel,
Institute for Physics, CH-4056 Basel, Switzerland}
and Y.~Bruynseraede}
\address{Laboratorium voor Vaste-Stoffysica en Magnetisme, Katholieke
Universiteit Leuven,\\ Celestijnenlaan 200 D, B-3001 Leuven, Belgium}

\maketitle

\begin{abstract}
We have studied quantization\index{quantization} and confinement\index{confinement} effects in
nanostructured superconductors. Three different types of nanostructured
samples were investigated: individual structures (line\index{mesoscopic line}, loop\index{mesoscopic loop}, dot\index{mesoscopic dot}),
1-dimensional (1D)~clusters of loops and 2D clusters of antidots, and finally large lattices
of antidots. Hereby, a crossover from individual elementary "plaquettes",
via clusters, to huge arrays of these elements, is realized.
The main idea of our study was to vary
the boundary conditions\index{boundary conditions} for confinement\index{confinement} of the superconducting condensate
by taking samples of different topology and, through that, modifying the
lowest Landau level\index{Landau levels} $E_{LLL}(H)$. Since the critical temperature versus
applied magnetic field $T_{c}(H)$ is, in fact, $E_{LLL}(H)$ measured in
temperature units, it is varied as well when the sample topology is changed 
through nanostructuring\index{nanostructuring}.
We demonstrate that in all studied nanostructured superconductors
the shape of the $T_{c}(H)$ phase boundary is determined by the confinement\index{confinement}
topology in a unique way.
\end{abstract}

\section{Introduction}
\subsection{Confinement\index{confinement} and Quantization\index{quantization}}
\begin{figure}[h]
\centerline{\psfig{figure=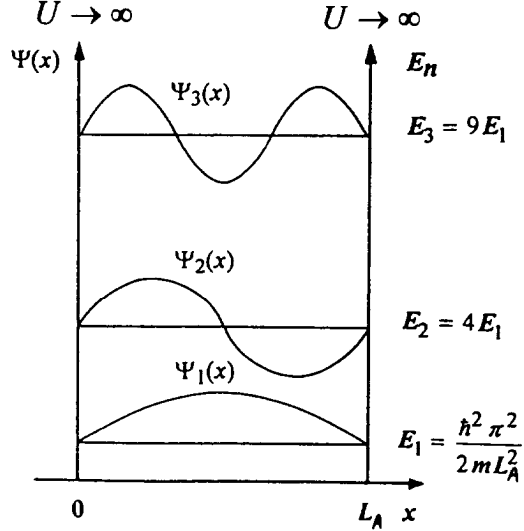}}
\caption{Confinement\index{confinement} and quantization\index{quantization} of the motion of a particle
by an infinite potential well with size $L_{A}$ for $n$ = 1, 2 and 3.}
\label{FA}
\end{figure}

"Confinement" and "quantization" are two closely interrelated definitions:
if a particle is "confined" then its energy is "quantized", and vice versa.
According to the
dictionary, to "confine" means to "restrict within limits", to "enclose",
and even to "imprison". A typical example, illustrating the relation
between confinement and quantization, is the restriction of the motion of
a particle by an infinite potential well of size $L_{A}$.  Due to the
presence of an infinite potential $U(x)$ (Fig.~\ref{FA}) for $x < 0$ and
$x > L_{A}$,
the wave function $\Psi(x)$ describing the particle is zero outside the
well: $\Psi = 0$ for $x < 0$ and $x > L_{A}$ and, in the region with
$U(x)=0 \: \: (0\leq x \leq L_{A})$, the solutions of
the one-dimensional Schr\"{o}dinger equation\index{Schr\"{o}dinger equation} correspond to standing
waves with an integer number $n$ of half wavelengths $\lambda$ along
$L_{A}: n \: {\lambda}_{n}/2 = L_{A}$.
This simple constraint results in the well-known quantized energy spectrum
\begin{equation}
E_{n}=\frac{\hbar^{2}k_{n}^{2}}{2m}=
\frac{\hbar^{2}(2\pi/\lambda_{n})^{2}}{2m}=
\frac{\hbar^{2}\pi^{2}}{2mL_{A}^{2}}n^{2}
\label{EnBox}
\end{equation}
Here $k_{n}$ is the wave number and $m$ is the free electron mass.
To have an idea about the characteristic energy scales involved and their
dependence upon the confinement length\index{confinement length} $L_{A}$, we have calculated (see 
Table~1) the energies $E_1$ (Eq.~(\ref{EnBox})) for electrons confined by
an infinite potential well with the sizes
$1$~\AA, $1 \: nm$ and $1 \: \mu m$.
\begin{center}
\begin{table}[h]
\caption{Confinement by the infinite potential well}
\label{table1}
\begin{tabular}{ccc}
\multicolumn{1}{c}{Confinement length $L_{A}$}& 
\multicolumn{1}{c}{Energy $E_1$}&
\multicolumn{1}{c}{Temperature $T$}\\
\tableline
$1$~\AA & $40 \: eV$ & $4 \times 10^5 \: K$ \\
$1 \: nm$ & $0.4 \: eV$ & $4 \times 10^3 \: K$ \\
$1 \: \mu m$ & $0.4 \: \mu eV$ & $4 \: mK$ \\
\end{tabular}
\end{table}
\end{center}
\subsection{Nanostructuring\index{nanostructuring}}
Recent impressive progress in nanofabrication has made it
possible to realize the whole range of confinement 
lengths $L_A$ : from $1 \: \mu m$ (photo-and e-beam lithography),
via $1 \: nm$ to $1$~\AA~(single atom manipulation) 
and, through that, to control the confinement\index{confinement} energy (temperature) from
a few $mK$ higher up to far above room temperature (Table~1).

This progress has stimulated dramatically the experimental and theoretical
studies of different nanostructured materials\index{nanostructured materials} and individual nanostructures\index{nanostructures}.
The interest towards such structures arises from the remarkable principle of 
"quantum design", when quantum mechanics can be efficiently used to tailor
the physical properties of nanostructured materials\index{nanostructured materials}.

\begin{figure}[h]
\centerline{\psfig{figure=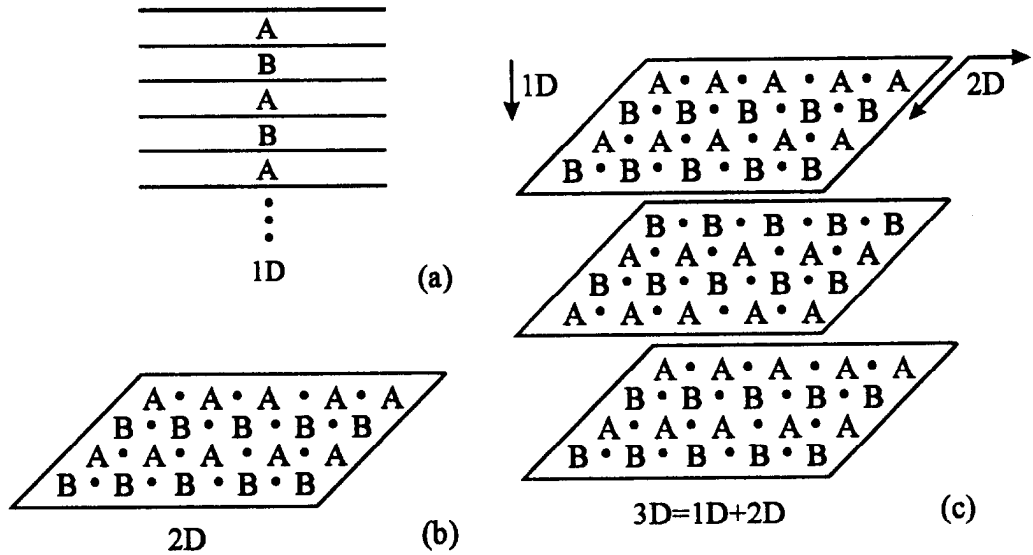}}
\caption{Schematic presentation of the vertical modulation in
superlattices or multilayers (a), of the horizontal modulation achieved by
a lateral repetition of elements A and B (b) and of the
1D+2D=3D artificial modulation~(c).}
\label{FB}
\end{figure}

Nanostructuring\index{nanostructuring} can also be considered  as a sort of artificial
modulation.  We can identify then the main classes of nanostructured 
materials using the idea of their modification along one-, two- or
three-axes, thus introducing 1-dimensional (1D)-, 2D- or 3D- artificial
modulation (Fig.~\ref{FB}).
The 1D or "vertical" modulation represents then the class of
superlattices or multilayers (Fig.~\ref{FB}a) formed by alternating
individual films of two (A, B) or more different materials on a stack.
Some examples of different types of multilayers are
superconductor/insulator (Pb/Ge, WGe/Ge,...), superconductor/metal
(V/Ag,...), superconductor/ferromagnet (Nb/Fe, V/Fe,...), ferromagnet/metal
(Fe/Cr, Cu/Co,...), etc.

The "horizontal" (lateral) superlattices (Fig.~\ref{FB}b) correspond to
the 2D artificial modulation achieved by a lateral repetition of one (A),
two (A,B) or more elements.  As examples, we should mention
here antidot arrays\index{antidot array} or antidot lattices\index{antidot lattice}, when A=microhole ("antidot"),
or arrays and lateral superlattices consisting of magnetic dots.

If the 2D lateral modulation is applied to each individual layer of
a multilayer or superlattice, then we deal with the 1D+2D=3D artificial
modulation (Fig.~\ref{FB}c). For example, if arrays of antidots\index{antidot array} are made in
a multilayer, then we have a system with 3D artificial modulation
which combines 2D lateral "horizontal" with the 1D "vertical" modulation.

\begin{figure}[h]
\centerline{\psfig{figure=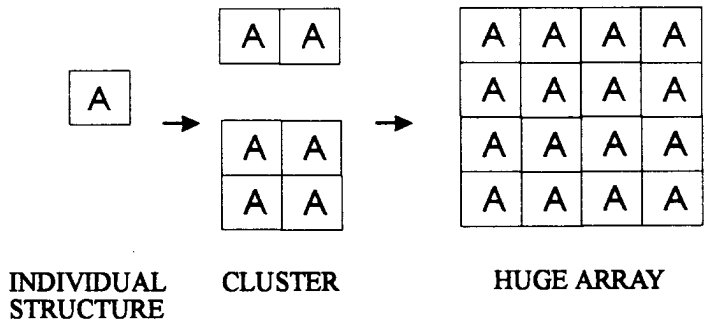}}
\caption{Schematic presentation of an individual structure, a cluster and
a huge array.}
\label{FC}
\end{figure}

Finally, macroscopic nanostructured samples, with a huge number $N$ of 
repeated elementary "plaquettes" (A,B,...), are examples of very complicated
systems if the confined charge carriers or flux lines are
strongly interacting with each other and the relevant interaction is of a
long range.  In this case the essential physics of such systems can be
understood much better if we use clusters of elements ($N\simeq$ 10),
instead of their huge arrays ($N\rightarrow\infty$) (Fig.~\ref{FC}).
These clusters, occupying an intermediate place between individual
nanostructures\index{nanostructures} ($N=1$) and nanostructured materials\index{nanostructured materials} ($N\rightarrow\infty$),
are very helpful model objects to study the interactions between flux lines
or charge carriers confined by elements A. The "growth" of clusters on the
way from an individual object A to a huge
array of A's can be done either in a 1D or 2D fashion (Fig.~\ref{FC}), thus
realizing 1D chains or 2D-like clusters of elements A.

\subsection{Confining the Superconducting Condensate}

The nanostructured materials\index{nanostructured materials} and individual nanostructures\index{nanostructures}, introduced
in the previous section, can be prepared using the modern facilities for
nanofabrication. It is worth, however, first asking ourselves
a few simple questions like: why do we want to make such structures,
what interesting new physics do we expect, and why do we want to focus on
superconducting (and not, for example, normal metallic) nanostructured
materials?

First of all, by making nanostructured materials\index{nanostructured materials}, one creates an
artificial potential\index{artificial potential} in which charge carriers or flux lines are confined.
The size $L_{A}$ of an elementary "plaquette" A, gives roughly the expected
energy scale in accordance with Table~1, while the positions
of the elements A determine the pattern of the potential modulation.
The concentration of charge carriers or flux lines can be controlled by
varying a gate voltage (in 2D electron gas systems)\cite{Ensslin} or the applied
magnetic field (in superconductors)\cite{Pannetier84}. In this situation,
different commensurability\index{commensurability} effects
between the fixed number of elements A in an array and a tunable number
of charge or flux carriers are observed.

Secondly, modifying the sample topology in nanostructured
materials creates a unique possibility to impose the desired boundary
conditions, and thus almost "impose" the properties of the sample.
A Fermi liquid or a
superconducting condensate confined within such materials will be subjected
to severe constraints and, as a result, the properties of these materials
will be strongly affected by the boundary conditions\index{boundary conditions}.

While a normal metallic system should be considered quantum-mechanically
by solving the Schr\"{o}dinger equation\index{Schr\"{o}dinger equation}:
\begin{equation}
\frac{1}{2 m}
\left( - \imath \hbar \vec{\nabla} - e \vec{A} \right)^{2} \Psi +U \: \Psi
= E \: \Psi \: ,
\label{e}
\end{equation}
a superconducting system is described by the two coupled
Ginzburg-Landau (GL) equations\index{Ginzburg-Landau equations}:
\begin{equation}
\frac{1}{2m^{\star}}(-i \hbar\vec{\nabla}-e^{\star} \vec{A})^{2}\Psi_s+
\beta |\Psi_s|^{2} \Psi_s = -\alpha \Psi_s
\label{GLFree1}
\end{equation}
\begin{equation}
\vec{j_{S}}=\vec{\nabla} \times \vec{h} = \frac{e^{\star}}{2 m^{\star}} 
\left[ \Psi_s^{\star} (- \imath \hbar \vec{\nabla} - e^{\star} \vec{A})
\Psi_s +
\Psi_s ( \imath \hbar \vec{\nabla} - e^{\star} \vec{A}) \Psi_s^{\star}
\right] \: ,
\label{GL2A}
\end{equation}
with $\vec{A}$ the vector potential which corresponds to the microscopic
field $\vec{h}=rot\vec{A}/\mu_{0}$, $U$ the potential energy, $E$ the total
energy, $\alpha$ a temperature dependent parameter changing sign from 
$\alpha > 0$ to $\alpha < 0$ as $T$ is decreased through
$T_{c}$, ${\beta}$ a positive temperature independent constant,
$m^{\star}$ the effective mass which can be chosen arbitrarily
and is generally taken as twice the free electron mass $m$.

Note that the first GL~equation\index{Ginzburg-Landau equations} (Eq.~(\ref{GLFree1})), with the nonlinear
term $\beta |\Psi_s|^{2} \Psi_s$ neglected, is the analogue of the
Schr\"{o}dinger equation\index{Schr\"{o}dinger equation} (Eq.~(\ref{e})) with $U$ = 0, when making a few
substitutions: $\Psi_{s}\leftrightarrow\Psi$, $e^{\star}\leftrightarrow e$,
$-\alpha\leftrightarrow E$ and $m^{\star} \leftrightarrow m$.
The superconducting order parameter
$\Psi_{s}$ corresponds to the wave function $\Psi$ in Eq.~(\ref{e}).
The effective charge $e^{\star}$ in the GL equations\index{Ginzburg-Landau equations} is $2 e$, i.e.
the charge of a Cooper pair, while the temperature dependent GL parameter
$\alpha$
\begin{equation}
- \alpha = \frac{\hbar^{2}}{2 m^{\star} \: \xi^{2}(T)}
\label{GLAlpha}
\end{equation}
plays the role of $E$ in Schr\"{o}dinger equation\index{Schr\"{o}dinger equation}.  Here $\xi(T)$ is the
temperature dependent coherence length\index{coherence length}:
\begin{equation}
\xi(T)=\frac{\xi(0)}{\sqrt{1-\frac{T}{T_{c0}}}}.
\label{XiT}
\end{equation}

\begin{figure}[h]
\centerline{\psfig{figure=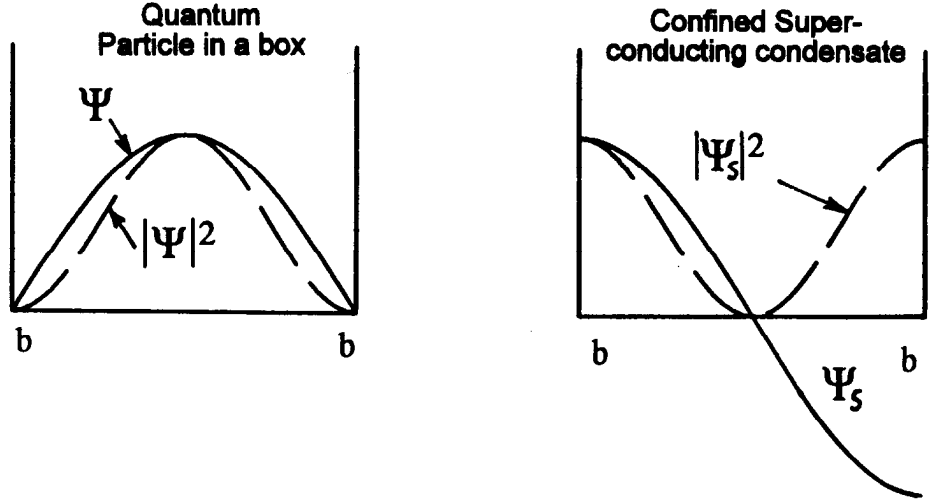}}
\caption{Boundary conditions for interfaces normal metal-vacuum and
\mbox{superconductor}-vacuum.}
\label{FD}
\end{figure}

The boundary conditions\index{boundary conditions} for interfaces normal metal-vacuum and
superconductor-vacuum are, however, different (Fig.~\ref{FD}):
\begin{equation}
\left. \Psi\Psi^{\star} \right|_{b}=0
\label{EBound}
\end{equation}
\begin{equation}
\left. (- \imath \hbar \vec{\nabla} - e^{\star} \vec{A}) \Psi_{s}
\right|_{\perp , b}=0
\label{GLBound}
\end{equation}
i.e. for normal metallic systems the density is zero, while for
superconducting systems, the gradient of $\Psi_s$ (for the case $\vec{A}=0$) 
has no component perpendicular to the boundary. As a consequence, the
supercurrent\index{supercurrent} cannot flow through the boundary. The nucleation of the 
superconducting condensate is favored at the superconductor/
vacuum interfaces, thus leading to the appearance of superconductivity
in a surface\index{surface superconductivity} sheet with a thickness $\xi(T)$ at the third critical field
$H_{c3}(T)$.

For bulk superconductors the surface-to-volume ratio is negligible and
therefore superconductivity in the
bulk is not affected by a thin superconducting surface layer.
For nanostructured superconductors with antidot arrays\index{antidot array}, however, 
the boundary conditions\index{boundary conditions} (Eq.~(\ref{GLBound})) and the surface superconductivity\index{surface superconductivity}
introduced through them, become very important if $L_{A}\leq\xi(T)$.  The
advantage of superconducting materials in this case is that it is not even
necessary to go to $nm$ scale (like for normal metals), since for $L_{A}$ of the order of
0.1-1.0~$\mu m$ the temperature range where $L_{A} \leq \xi(T)$, spreads
over $0.01-0.1~K$ below $T_{c}$ due to the divergence of $\xi(T)$ at
$T\rightarrow T_{c0}$ (Eq.~(\ref{XiT})).

In principle, the mesoscopic regime $L_{A}\leq\xi(T)$ can 
be reached even in bulk superconducting samples with $L_{A}\sim$
$1~cm -1~m$, since $\xi(T)$ diverges. However, the temperature window where
$L_{A}\leq\xi(T)$ is so narrow, not more than $\sim$ $1~nK$
below $T_{c0}$, that one needs ideal sample homogeneity and
perfect temperature stability.

In the mesoscopic regime $L_{A}\leq\xi(T)$, which is quite easily realized
in (perforated) nanostructured materials\index{nanostructured materials}, the surface superconductivity\index{surface superconductivity} can
cover the whole available space occupied by the material, thus spreading
superconductivity all over the sample.  It is then evident that in this
case surface effects play the role of bulk effects.

Using the similarity between the linearized GL~equation\index{Ginzburg-Landau equations}
(Eq.~(\ref{GLFree1})) and the Schr\"{o}dinger equation\index{Schr\"{o}dinger equation} (Eq.~(\ref{e})),
we can formalize our approach as follows:  since the parameter
-$\alpha$ (Eqs.~(\ref{GLFree1}) and~(\ref{GLAlpha})) plays the role of
energy $E$ (Eq.~(\ref{e})), then {\it the highest possible temperature
$T_{c}(H)$ for the nucleation of the
super\-con\-duc\-ting state in presence of the magnetic field $H$ always
corresponds to the lowest Landau level\index{Landau levels} $E_{LLL}(H)$} found by solving the
Schr\"{o}dinger equation\index{Schr\"{o}dinger equation} (Eq.~(\ref{e})) with "superconducting" boundary
conditions\index{boundary conditions}~(Eq.~(\ref{GLBound})).

\begin{figure}[h]
\centerline{\psfig{figure=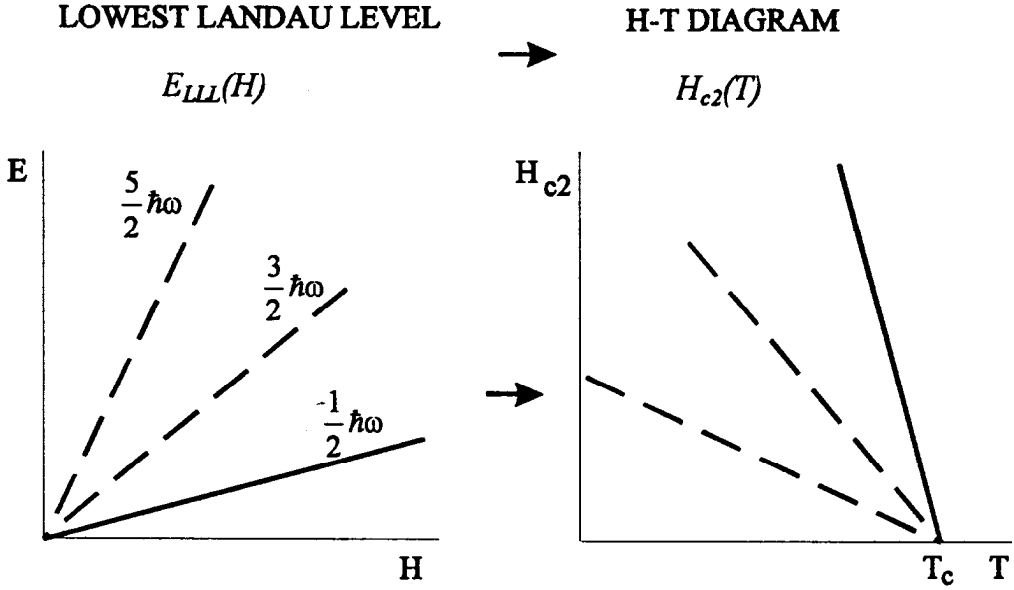}}
\caption{Landau level scheme for a particle in a magnetic field. From the
lowest Landau level $E_{LLL}(H)$ the second critical field $H_{c2}(T)$
is derived (solid line).}
\label{FE}
\end{figure}
Figure~\ref{FE} illustrates the application of this rule to the calculation
of the upper critical field $H_{c2}(T)$: indeed, if we take the well-known
classical Landau solution for the lowest level in bulk samples
$E_{LLL}(H)=\hbar\omega/2$, where $\omega = e^{\star} \mu_0 H / m^{\star}$
is the cyclotron frequency. Then, from -$\alpha = E_{LLL}(H)$ we have 
\begin{equation}
\frac{\hbar^{2}}{2 m^{\star} \: \xi^{2}(T)}=
\left. \frac{\hbar\omega}{2}\right|_{H=H_{c2}}
\label{ha}
\end{equation}
and with the help of Eq.~(\ref{GLAlpha}), we obtain
\begin{equation}
\mu_{0} H_{c2}(T)=\frac{\Phi_{0}}{2 \pi \xi^{2}(T)}
\label{hc2}
\end{equation}
with $\Phi_0 = h/e^{\star} = h/2 e$ the superconducting flux quantum.

In nanostructured superconductors, where the boundary conditions\index{boundary conditions}
(Eq.~(\ref{GLBound})) strongly influence the Landau level\index{Landau levels} scheme,
$E_{LLL}(H)$ has to be calculated for each different confinement\index{confinement} geometry.
By measuring the shift of the critical temperature $T_c(H)$ in a magnetic
field, we can compare the experimental 
$T_{c}(H)$ with the calculated level $E_{LLL}(H)$ and thus check
the effect of the confinement\index{confinement} topology on the superconducting phase
boundary for a series of nanostructured superconducting samples. The
transition between normal and superconducting states is usually very sharp
and therefore the lowest Landau level\index{Landau levels} can be easily traced as a function of
applied magnetic field. Except when stated explicitly, we have taken the
midpoint of the resistive transition from the superconducting to the normal
state, as the criterion to determine $T_c(H)$.

In this paper we present the systematic study of the influence of the
confinement\index{confinement} geometry on the superconducting phase boundary $T_{c}(H)$ in a
series of nanostructured samples. We begin with individual
nanostructures\index{nanostructures} of different topologies (lines\index{mesoscopic line}, loops\index{mesoscopic loop}, dots\index{mesoscopic dot}) (Section II) and
then focus on "intermediate" systems: clusters of loops fabricated in the
form of a 1D chain of loops (Section III.A) or 2D antidot
clusters (Section III.B).
Finally, we move on to huge arrays of antidots\index{antidot array} in Section IV where we report
on the $T_{c}(H)$ boundary for superconducting films with antidot
lattices. The main emphasis of the paper is on the demonstration of the
importance of the confinement\index{confinement} geometry for the superconducting condensate
and on the related quantization\index{quantization} phenomena in nanostructured
superconductors through the study of the phase boundaries $T_{c}(H)$.

\section{Individual structures: line\index{mesoscopic line}, loop\index{mesoscopic loop} and dot\index{mesoscopic dot}}
We begin this section by presenting the experimental results on the 
$T_c(H)$ phase boundary of individual superconducting mesoscopic structures
of different topology. Simultaneously, we have kept other parameters of
these samples, like material from which they are made (Al), the width of
the lines ($w=0.15 \: \mu m$) and the film thickness $t= 25 \: nm$ the same
for all three structures, thus directly relating the differences in
$T_c(H)$ to topological effects. The magnetic field $H$ is always applied
perpendicular to the structures.

\subsection{Line\index{mesoscopic line}}
\begin{figure}[H]
\centerline{\psfig{figure=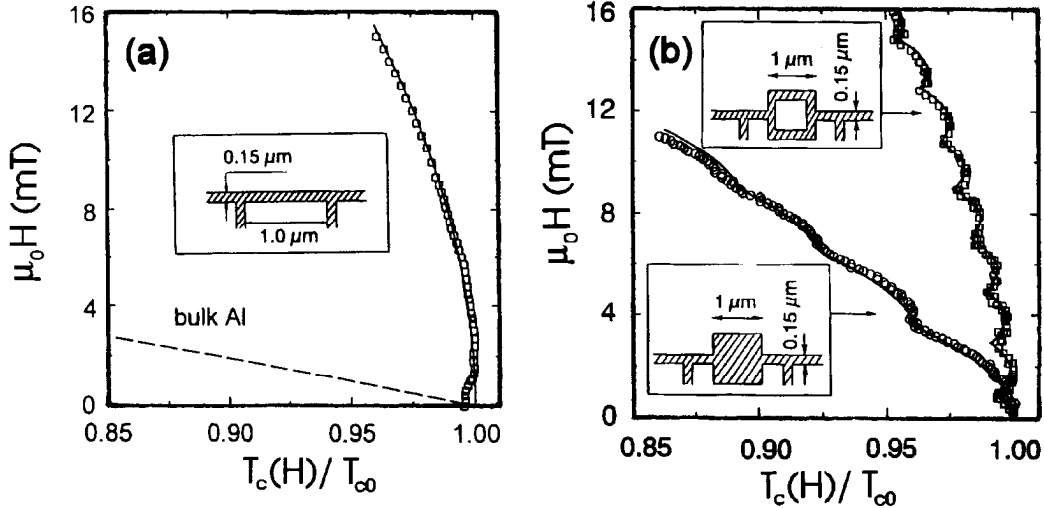}}
\caption{The measured superconducting/normal phase boundary as a function
of the reduced temperature $T_c(H)/T_{c0}$ for a)~the line, and b)~the loop
and the dot. The solid line in (a) is calculated using Eq.~(\ref{TCBLine})
with $\xi (0)=110 \: nm$ as a fitting parameter. The dashed line represents
$T_c(H)$ for bulk Al.}
\label{MesLine}
\end{figure}

In Fig.~\ref{MesLine}a the phase boundary $T_c(H)$ of a mesoscopic line\index{mesoscopic line}
is shown. The solid line gives the $T_c(H)$ calculated from the well-known
formula\cite{Tin63}:
\begin{equation}
T_{c}(H)=T_{c0} \left[ 1 - \frac{\pi^{2}}{3}
\left( \frac{w \: \xi(0) \mu_0 H}{\Phi_0}\right)^{2} \right]
\label{TCBLine}
\end{equation}
which, in fact, describes the parabolic shape of $T_c(H)$ for a thin film
of thickness $w$ in parallel magnetic field. Since the cross-section,
exposed to the applied magnetic field, is the same for a film of thickness
$w$ in a parallel magnetic field and for a mesoscopic line\index{mesoscopic line} of width $w$ in a
perpendicular field, the same
formula can be used for both\cite{VVM95}. Indeed, the solid line in
Fig~\ref{MesLine}a is a parabolic fit of the experimental data with
Eq.~(\ref{TCBLine}) where $\xi (0)=110 \: nm$ was obtained as a fitting
parameter.
The coherence length\index{coherence length} obtained using this method, coincides reasonably well
with the dirty limit value $\xi(0)= 0.85 (\xi_0 \ell)^{1/2}= 132 \: nm$
calculated from the known BCS coherence length\index{coherence length} $\xi_0=1600 \: nm$ for bulk
Al\cite{dGABook} and the mean free path $\ell = 15 \: nm$, estimated from
the normal state resistivity $\rho$ at $4.2 \: K$\cite{Rom82}.

We can use also another simple argument to explain the parabolic relation
$T_c(H) \propto H^2$: the expansion of the energy $E(H)$ in powers of $H$,
as given by the perturbation theory, is\cite{Wel38}:
\begin{equation}
E(H)=E_0+A_1 L H + A_2 S H^2 + \cdots
\label{Perturb}
\end{equation}
where $A_1$ and $A_2$ are constant coefficients, the first term $E_0$
represents the energy levels in zero field, the second term is the linear
field splitting with the orbital quantum number\index{orbital quantum number} $L$ and the third term
is the diamagnetic shift with $S$, being the area exposed to the applied
field.

Now, for the topology of the line\index{mesoscopic line} with a width $w$ much smaller than the
Larmor radius $r_H \gg w$, any orbital
motion is impossible due to the constraints imposed by the
boundaries onto the electrons inside the line\index{mesoscopic line}. Therefore, in this particular
case $L=0$ and $E(H)=E_0 + A_2 S H^2$, which immediately leads to the
parabolic relation $T_c \propto H^2$. This diamagnetic shift of $T_c(H)$
can be understood in terms of a partial screening of the magnetic field
$H$ due to the non-zero width of the line\cite{Tinkham75}.

\subsection{Loop\index{mesoscopic loop}}
The $T_c(H)$ of  the mesoscopic loop, shown in Fig.~\ref{MesLine}b, 
demonstrates very distinct Little-Parks\index{Little-Parks effect} (LP) oscillations\cite{Litt62}
superimposed on a monotonic background. A closer investigation
leads to the conclusion that this background is very well described by
the same parabolic dependence as the one which we just discussed for the
mesoscopic line\index{mesoscopic line}\cite{VVM95} (see the solid line in Fig.~\ref{MesLine}a).
As long as the width of the strips $w$, forming the loop, is much smaller
than the loop\index{mesoscopic loop} size, the total shift of $T_c(H)$ can be written as the
sum of an oscillatory part, and the monotonic background given by
Eq.~(\ref{TCBLine})\cite{VVM95,Gro68}:
\begin{equation}
T_c(H)=T_{c0} \left[ 1 - \frac{\pi^2}{3} \left( 
\frac{w \: \xi(0) \mu_0 H}{\Phi_0} \right)^2 - \frac{\xi^2(0)}{R^2}
\left( n - \frac{\Phi}{\Phi_0} \right)^2 \right]
\label{TCBRing}
\end{equation}
where $R^2=R_1 \: R_2$ is the product of inner and outer loop radius, and
the magnetic flux threading the loop $\Phi=\pi R^2 \mu_0 H$. The integer
$n$ has to be chosen so as to maximize $T_c(H)$ or, in other words,
selecting $E_{LLL}(H)$.

The LP~oscillations\index{Little-Parks effect} originate from the fluxoid quantization\index{fluxoid quantization} requirement,
which states that the complex order parameter $\Psi_s$ should be a
single-valued function when integrating along a closed contour
\begin{equation}
\oint \vec{\nabla} \varphi \cdot dl = n \: 2 \pi 
\; \; \; \; \; \; \; \; \; \; \; \; \; \; \; \;
n=\cdots \: , -2,-1,0,1,2,  \: \cdots
\label{Fluxoid}
\end{equation}
where we have introduced the order parameter as
$\Psi_s = |\Psi_s| \exp \: (\imath \varphi)$. Fluxoid quantization\index{fluxoid quantization} gives
rise to a circulating supercurrent\index{supercurrent} in the loop\index{mesoscopic loop} when $\Phi \neq n \Phi_0$,
which is periodic with the applied flux $\Phi / \Phi_0$.

Using the sample dimensions and the value for $\xi (0)$ obtained before
for the mesoscopic line\index{mesoscopic line} (with the same width $w=0.15 \: \mu m$), the
$T_c(H)$ for the loop can be calculated from Eq.~(\ref{TCBRing}) without
any free parameter. The solid line in Fig.~\ref{MesLine}b shows indeed
a very good agreement with the experimental data.
It is worth noting here that the amplitude of the
LP~oscillations\index{Little-Parks effect} is about a few $mK$ - in qualitative agreement with the
simple estimate given in Table~1 for $L_A \simeq 1 \: \mu m$.

Another interesting feature of a mesoscopic loop\index{mesoscopic loop} and
other mesoscopic structures is the unique possibility they offer for
studying nonlocal effects\index{nonlocal effects}\cite{StrN96}.
In fact, a single loop can be considered as a
2D artificial quantum orbit with a {\it fixed radius}, in contrast to Bohr's
description of atomic orbitals. In the latter case the
stable radii are found from the quasiclassical quantization\index{quantization} rule,
stating that only an integer number of wavelengths
can be set along the circumference of the allowed orbits. For a
superconducting loop\index{mesoscopic loop}, however, supercurrents\index{supercurrent} must flow, in order to
fulfil the fluxoid quantization\index{fluxoid quantization} requirement (Eq.~(\ref{Fluxoid})), thus
causing oscillations of the critical temperature $T_c$ versus magnetic
field $H$.

In order to measure the resistance of a mesoscopic loop\index{mesoscopic loop}, electrical
contacts have, of course, to be attached to it, and as a consequence
the confinement\index{confinement} geometry is changed. A loop\index{mesoscopic loop} with attached contacts and
the same loop without any contacts are, strictly speaking, different
mesoscopic systems. This "disturbing" or "invasive" aspect can now be
exploited for the study of nonlocal effects\index{nonlocal effects}\cite{StrN96}.
Due to the divergence of the
coherence length\index{coherence length} $\xi(T)$ at $T = T_{c0}$ (Eq.~(\ref{XiT}))
the coupling of the loop with the attached leads is expected to be very
strong for $T \rightarrow T_{c0}$.

\begin{figure}[h]
\centerline{\psfig{figure=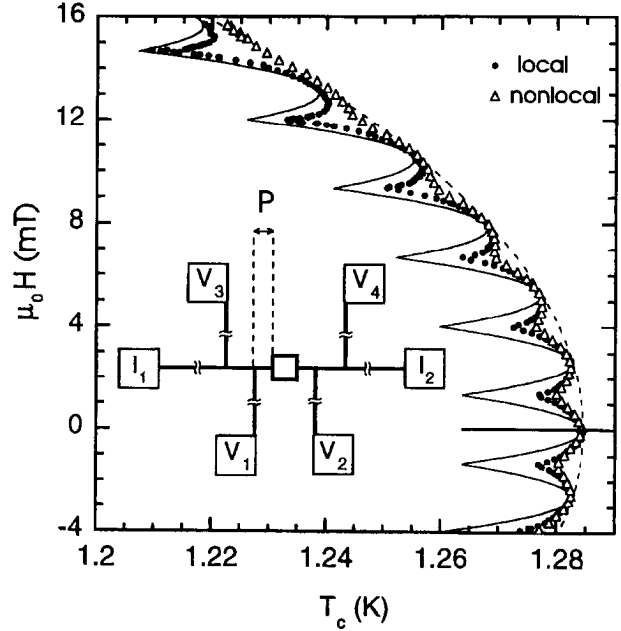}}
\caption{Local ($V_1/V_2$) and nonlocal phase boundaries ($V_1/V_3$
or $V_2/V_4$) $T_c(H)$. The mesuring current is sent through $I_1/I_2$.
The solid and dashed lines correspond to the theoretical $T_c(H)$ of
an isolated loop and a one-dimensional line\index{mesoscopic line}, respectively. The inset shows
a schematic of the structure, where the distance $P=0.4 \: \mu m$.}
\label{NL1}
\end{figure}

Fig.~\ref{NL1} shows the results of these measurements. Both "local"
(potential probes across the loop $V_1/V_2$)
and "nonlocal" (potential probes aside of the loop $V_1/V_3$
or $V_2/V_4$) LP~oscillations\index{Little-Parks effect} are clearly observed. For the "local" probes
there is an unexpected and pronounced increase of the oscillation amplitude
with increasing field, in disagreement with previous measurements on Al
microcylinders\cite{Gro68}. In contrast to this, for the "nonlocal"
LP~effect, the oscillations rapidly vanish when the magnetic field is
increased.

When increasing the field, the background suppression of $T_c$
(Eq.~(\ref{TCBLine})) results in a decrease of $\xi(T)$. Hence, the change
of the oscillation amplitude with $H$ is directly related to the
temperature-dependent coherence length\index{coherence length}. As long as the coherence of
the superconducting condensate protuberates over the nonlocal\index{nonlocal effects} voltage
probes, the nonlocal LP~oscillations\index{Little-Parks effect} can be observed.

On the other hand, the importance of an "arm" attached to a mesoscopic
loop\index{mesoscopic loop}, was already demonstrated theoretically by de Gennes in
1981\cite{dGA81}. For a perfect 1D loop (vanishing width of the
strips) adding an "arm" will result in a decrease of the LP~oscillation
amplitude, what we observed indeed at low magnetic fields, where $\xi(T)$
is still large. With these experiments, we have proved that adding probes
to a structure considerably changes both the confinement\index{confinement} topology and the phase boundary $T_c(H)$.

\subsection{Dot\index{mesoscopic dot}}

\begin{figure}[h]
\centerline{\psfig{figure=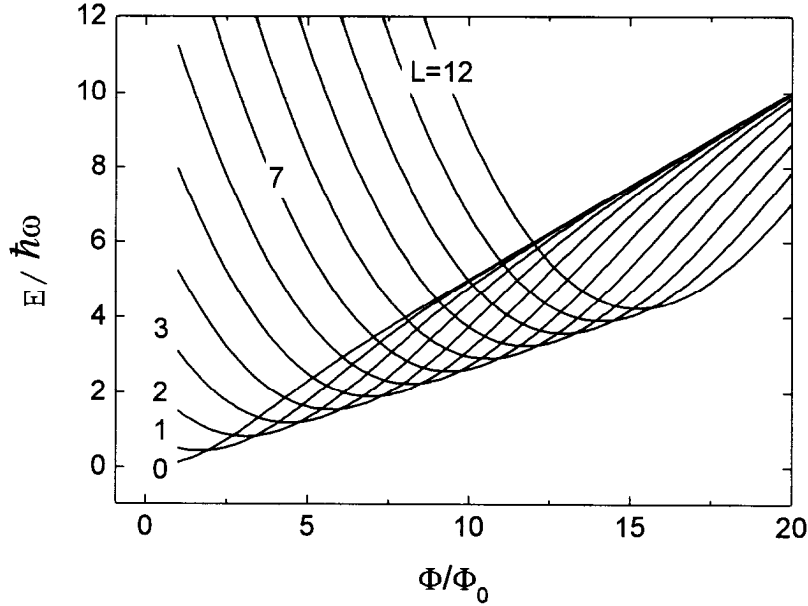}}
\caption{Energy level scheme versus normalized flux $\Phi / \Phi_0$
for a superconducting cylinder in a parallel magnetic field.
The cusp-like $H_{c3}(T)$ line is formed due to the change of the orbital
quantum number $L$.}
\label{DotCalc}
\end{figure}

The Landau level\index{Landau levels} scheme for a cylindrical dot with "superconducting"
boundary conditions\index{boundary conditions} (Eq.~(\ref{GLBound})) is presented in
Fig.~\ref{DotCalc}. Each level is characterized by a certain orbital
quantum number $L$ where
$\Psi_s = |\Psi_s| \exp \: (\mp \imath L \varphi)$\cite{PME96}.
The levels, corresponding
to the sign "+" in the argument of the exponent are not shown since they
are situated at energies higher than the ones with the sign "-". 
The lowest Landau level\index{Landau levels} in Fig.~\ref{DotCalc} represents a cusp-like
envelope, switching between different $L$ values with changing magnetic field.
Following our main guideline that $E_{LLL}(H)$ determines $T_c(H)$, we
expect for the dot\index{mesoscopic dot} the cusp-like superconducting phase boundary\index{superconducting phase boundary} with
nearly perfect linear background. The measured phase boundary $T_c(H)$,
shown in Fig.~\ref{MesLine}b,
can be nicely fitted by the calculated one (Fig.~\ref{DotCalc}), 
thus proving that $T_c(H)$ of a superconducting dot\index{mesoscopic dot} indeed consists of
cusps with different $L$'s\cite{Bui90}. Each fixed $L$ describes a
giant vortex state which carries $L$ flux quanta $\Phi_0$. The linear background
of the $T_c(H)$ dependence is very close to the third critical field
$H_{c3}(T) \simeq 1.69 \: H_{c2}(T)$\cite{Saint-James65}.
Contrary to the loop, where the LP~oscillations\index{Little-Parks effect} are perfectly periodic,
the dot\index{mesoscopic dot} demonstrates a certain aperiodicity\cite{VVMScripta},
in very good agreement
with the theoretical calculations\cite{Bui90,Benoist}.

\section{Clusters of loops and antidots}
\subsection{1D Clusters of loops}

\begin{figure}[h]
\centerline{\psfig{figure=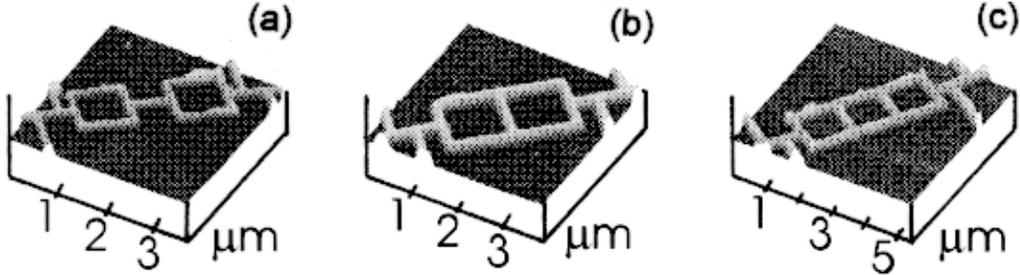}}
\caption{AFM micrographs of the studied structures: a)~the bola,
b)~the double loop, and c)~the triple loop.}
\label{afmsien}
\end{figure}
After the description of the confinement\index{confinement} effects of several individual
superconducting structures (A = line, loop, dot) we are ready to move
further on to clusters of elements A (Fig.~\ref{FC}) on our way
from "single plaquette" samples to materials nanostructured by
introducing huge arrays of elements A. First we take A = loop and
consider one-dimensional multiloop structures\index{multiloop structures}: "bola", double and
triple loop Al structures. Figure~\ref{afmsien} shows
a AFM image of the structures. For these geometries
some interesting theoretical predictions have been made, for which no
experimental verification has been carried out up to know (see more
in Ref.\cite{Bru96}). The loops in all three
structures have the same dimensions, thus leading to the same magnetic field
period $\mu_0 \Delta H=1.24 \: mT$. The strips forming the structures
are $0.13 \: \mu m$ wide and the film thickness $t=34 \: nm$. In all
the experimental data we show, the parabolic background
(Eq.~(\ref{TCBLine})) is already subtracted in order to allow for a direct
comparison with the theory.
In the temperature interval where we measured
the $T_c(H)$ boundary\index{superconducting phase boundary}, the coherence length\index{coherence length} $\xi (T)$ is considerably
larger than the width $w$ of the strips.
This makes it possible to use the one-dimensional models for the 
calculation of $E_{LLL}(H)$ and thus $T_c(H)$. The basic
idea is to consider $|\Psi_s|$ = constant across the strips forming the
network and to allow a variation of $|\Psi_s|$ only along the strips.
In the simplest approach $|\Psi_s|$ is assumed to be spatially constant
(London limit\index{London limit} (LL))\cite{Halevi83,Chi92},
in contrast to the de~Gennes-Alexander (dGA)\index{de Gennes-Alexander approach}
approach\cite{dGA81,Alex83,FinkBola}, where $|\Psi_s|$ is allowed
to vary along the strips.
In the latter approach one imposes:
\begin{equation}
\sum_{n} \left( \imath \frac{\partial}{\partial x} + 
\frac{2 \pi}{\Phi_{0}} A_{\parallel}(x) \right) \Psi_{s}(x)=0
\label{GLCC}
\end{equation}
at the points where the current paths join.
The summation is taken over all strips connected to the junction point.
Here, $x$ is the coordinate defining the position on the strips, and
$A_{\parallel}$ is the component of the vector potential along $x$.
Eq.~(\ref{GLCC}) is often called the generalized first Kirchhoff law,
ensuring current conservation\cite{FinkBola}. The second Kirchhoff law
for voltages in normal circuits is now
replaced by the fluxoid quantization\index{fluxoid quantization} requirement (Eq.~(\ref{Fluxoid})),
which should be fulfilled for each closed contour around a loop.

\begin{figure}[H]
\centerline{\psfig{figure=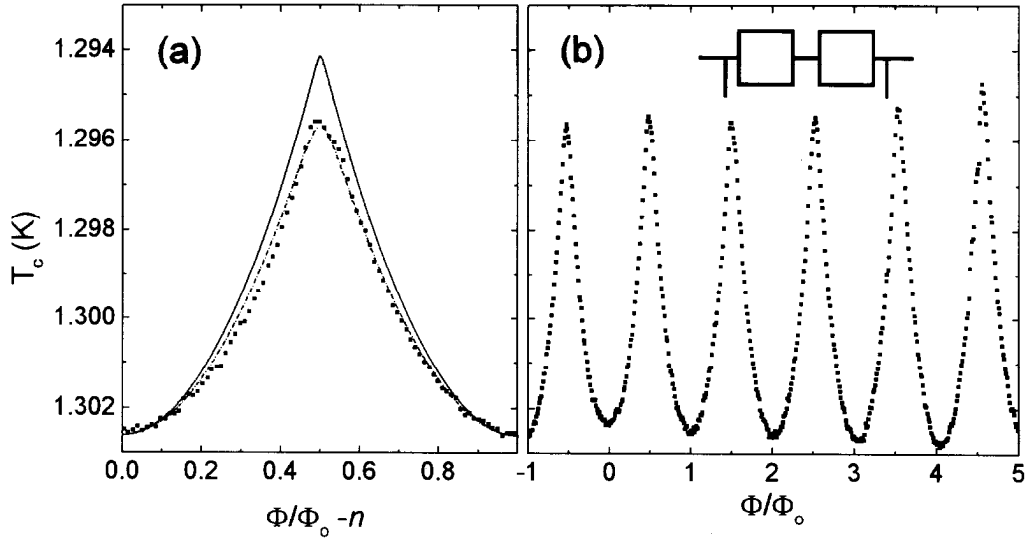}}
\caption{Experimental $T_c(\Phi)$ data for the {\bf bola} with the parabolic
background (Eq.~(\ref{TCBLine})) subtracted. The dots are the
experimental data points, while the lines correspond to the different
theoretical results as explained in the text. a)~Single period of 
$T_c(\Phi)$, b)~A few periods of the experimental $T_c(\Phi)$ curve.}
\label{bolsien}
\end{figure}

\begin{figure}[H]
\centerline{\psfig{figure=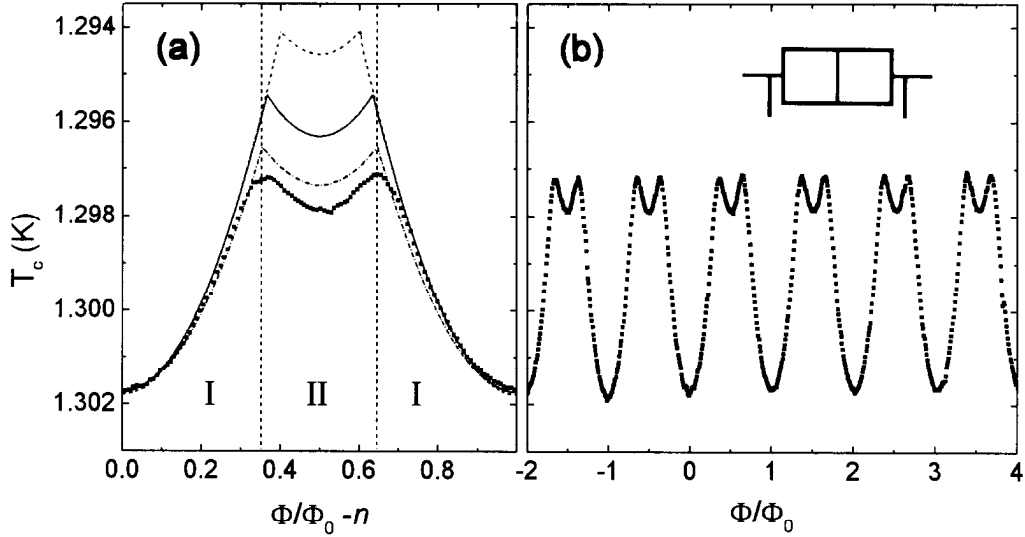}}
\caption{Experimental $T_c(\Phi)$ data for the {\bf double loop} with the
parabolic
background (Eq.~(\ref{TCBLine})) subtracted. The dots are the
experimental data points, while the lines correspond to the different
theoretical results as explained in the text. a)~Single period of 
$T_c(\Phi)$, b)~A few periods of the experimental $T_c(\Phi)$ curve.}
\label{doubsien}
\end{figure}

\begin{figure}[h]
\centerline{\psfig{figure=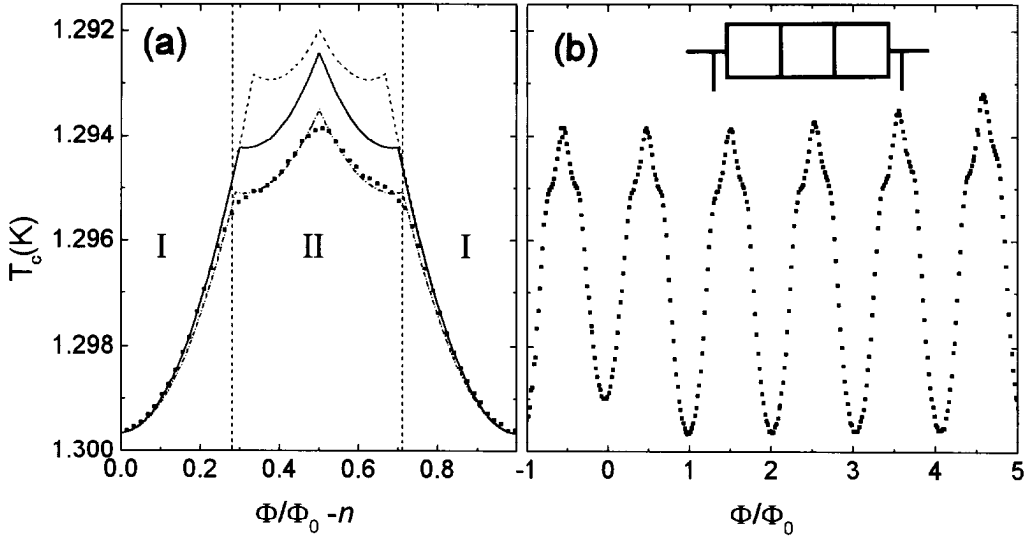}}
\caption{Experimental $T_c(\Phi)$ data for the {\bf triple loop} with the
parabolic
background (Eq.~(\ref{TCBLine})) subtracted. The dots are the
experimental data points, while the lines correspond to the different
theoretical results as explained in the text. a)~Single period of 
$T_c(\Phi)$, b)~A few periods of the experimental $T_c(\Phi)$ curve.}
\label{tripsien}
\end{figure}

In Figs.~\ref{bolsien}-\ref{tripsien} the phase boundaries of the
three structures are shown. The dashed lines are the phase
boundaries  obtained with the LL\index{London limit}, while the solid lines give the results
from the dGA approach\index{de Gennes-Alexander approach}. As we discussed in Section~II.B for a mesoscopic
loop, attaching contacts modifies the confinement\index{confinement} topology, so that the
amplitude of the local LP~oscillations\index{Little-Parks effect} is lowered at low magnetic fields.
Here as well, the inclusion of the leads reduces the amplitude of the
oscillations. The dash-dotted line in Figs.~\ref{bolsien}-\ref{tripsien}
give the result of the dGA\index{de Gennes-Alexander approach} calculation where the presence of the leads
has been included. The values for $\xi(0)$ obtained from the fits agree
within a few percent with the $\xi(0)$ values found independently from the
monotonic background of $T_c(\Phi)$ (see Eq.~(\ref{TCBLine})).

First, in Fig.~\ref{bolsien}, we consider the mesoscopic "bola" - two loops 
connected by a wire. Fink {\it et al.}\cite{FinkBola} showed that,
in the complete magnetic flux interval, the
spatially symmetric solution, with equal orientation of the supercurrents\index{supercurrent}
in both loops, has a lower energy than the antisymmetric solution.
Coming back to the similarity between a mesoscopic loop and a hydrogen
atom, we discussed in Section~II.B, we can then compare the bola with
a $H_2$ molecule, where the symmetric and the antisymmetric solutions
correspond to singlet and triplet states, respectively. 
In fact, $T_c(\Phi)$ of the bola is the
same as for a single loop provided that the length of the strip connecting
the two loops is short, as confirmed by the LP~oscillations\index{Little-Parks effect} observed
in the experimental $T_c(\Phi)$ (Fig.~\ref{bolsien}).

In what follows we will focus on the phase boundaries of the double
(Fig.~\ref{doubsien}) and the triple loop (Fig.~\ref{tripsien}).
To facilitate the discussion we divide the flux period in two intervals:
flux regime~I for $\Phi/\Phi_0 < { \it g}$ or $\Phi/\Phi_0 > (1-{ \it g})$
and flux regime~II for ${ \it g}< \Phi /\Phi_0 < (1-{ \it g})$.
In the flux regime~I the phase boundaries, predicted by
the different models, are nearly
identical. Near $\Phi/\Phi_0 = 1/2$ (flux regime~II),
however, clear differences are found between the dGA\index{de Gennes-Alexander approach} approach
and the LL\index{London limit}. The dGA\index{de Gennes-Alexander approach} result fits better the experimental data with 
respect to the crossover point {\it g} between regimes~I and~II, and the
amplitude of the $T_c$ oscillations.
Using the dGA\index{de Gennes-Alexander approach} approach we have calculated the spatial modulation
of $|\Psi_s|$ and the supercurrents\index{supercurrent} for different values at the
$T_c(\Phi)$ boundary.
In the flux regime~I $|\Psi_s|$ varies only slightly and therefore the 
results of the LL\index{London limit} and the dGA\index{de Gennes-Alexander approach} models nearly coincide. The 
elementary loops have an equal fluxoid quantum number (and 
consequently an equal supercurrent\index{supercurrent} orientation) for both the
double and the triple loop geometry. For the double loop this
leads to a cancellation of the supercurrent\index{supercurrent} in the middle
strip, while for the triple loop structure the fluxoid
quantization\index{fluxoid quantization} condition (Eq.~(\ref{Fluxoid})) results in a different value for the
supercurrent\index{supercurrent} in the inner and the outer loops. As a result,
the common strips of the triple loop structure carry a finite current.

\begin{figure}[h]
\centerline{\psfig{figure=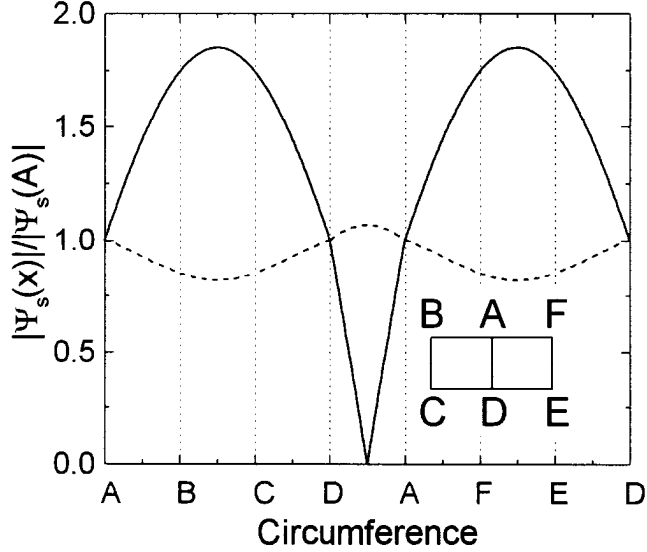}}
\caption{Calculated variation of $|\Psi_s|$ along the circumference of the
double loop, at the phase boundary ($\Phi / \Phi_0=0.36$). The dashed line
is the solution with $|\Psi_s|$ nearly
spatially constant, while the solid line is the state with a node in the
center of the strip connecting points $A$ and $D$.}
\label{orddpro}
\end{figure}

In the flux regime~II, qualitatively different states are obtained from the
LL\index{London limit} and the dGA\index{de Gennes-Alexander approach} approach: the states calculated within the dGA\index{de Gennes-Alexander approach} approach have
strongly modulated $|\Psi_s|$ along the strips. This is most severe for
the double loop: $\Psi_s$ shows a node ($|\Psi_s|=0$) in the center of
the common strip, the phase $\varphi$
having a discontinuity of $\pi$ at this point. This node is a
one-dimensional analog of the core of an Abrikosov vortex, where the order
parameter also vanishes and the phase shows a discontinuity.
In Fig.~\ref{orddpro} the spatial variation of $|\Psi_s|$ along the strips
is shown for $\Phi / \Phi_0=0.36$ close to the crossover point {\it g}.
The dashed curve gives $|\Psi_s|$ in flux regime~I, which is
quasi-constant. The strongly modulated solution, which goes through zero
in the center, is indicated by the solid line. Although there exists a
finite phase difference across the junction points of the middle strip,
no supercurrent\index{supercurrent} can flow through the strip due to the presence of the
node. This node is predicted to persist when moving below the phase
boundary\index{superconducting phase boundary} into the superconducting state\cite{Amm95,Cas95}.
Already in 1964 Parks\cite{Par64}
anticipated that, in a double loop, "a part of the middle link will revert
to the normal phase", and that "this in effect will convert the double
loop to a single loop", giving an intuitive explanation for the maximum in
$T_c(\Phi)$ at $\Phi / \Phi_0 =1/2$. Such a modulation of $|\Psi_s|$ is
obviously excluded in the LL\index{London limit}, 
where the loop currents have an opposite orientation and add up in the
central strip, thus giving rise to a rather high kinetic energy.
An extra argument in favor of the presence of the node is given by the
much better agreement for the crossover point {\it g} when the presence
of the leads is taken into account in the calculations
(see dash-dotted line in Fig.~\ref{doubsien}). 

For the triple loop structure the modulation of $|\Psi_s|$ is still
considerable 
in flux regime~II, but it does not show any nodes. Therefore the 
supercurrent\index{supercurrent} orientations can be found from the fluxoid quantum numbers
$\{n_i\}$, obtained from integrating the phase gradients along each
individual loop. When passing through the crossover point
between flux regime~I and regime~II only the supercurrent\index{supercurrent} in the middle loop is
reversed, while increasing the flux above $\Phi / \Phi_0 =1/2$
implies a reversal of the supercurrent\index{supercurrent} in all loops.

Surprisingly, the behavior of a microladder\index{microladder} with a linear arrangement of
$m$ loops appears to be qualitatively different for even and for odd $m$ in 
the sense that $m$ determines the presence or absence of nodes in the common
strips. For an infinitely long microladder\index{microladder} $|\Psi_s|$ was found to be
spatially constant below a certain $\Phi < \Phi_c$\cite{Sim82},
which is analogous to the
states we find in flux regime~I. For fluxes $\Phi > \Phi_c$ modulated
$|\Psi_s|$ states, with an incommensurate\index{commensurability} fluxoid pattern, were found.
At $\Phi / \Phi_0 =1/2$, nodes appear at the center of every second
common (transverse) branch.

\subsection{2D clusters of antidots}

As a 2D intermediate structure between individual elements A
and their huge arrays (Fig.~\ref{FC}), we have studied the
superconducting micro-square with a 2$\times$2 antidot cluster\index{antidot cluster}. In this
case A= "antidot". 
\begin{figure}[h]
\centerline{\psfig{figure=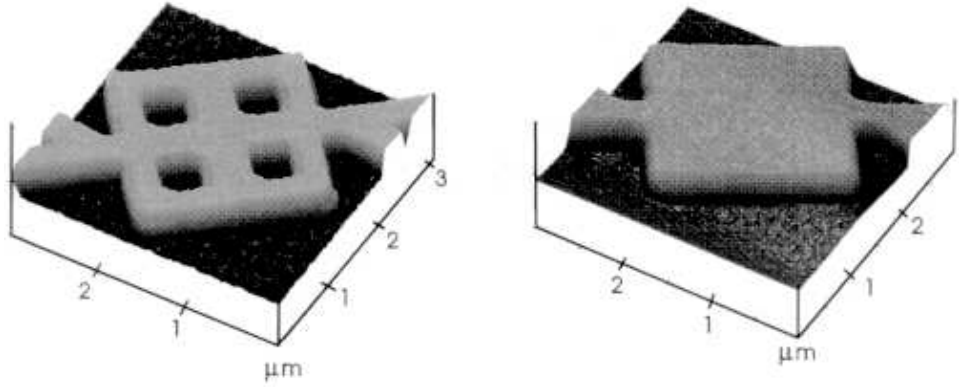}}
\caption{AFM image of the Pb/Cu 2$\times$2 antidot cluster\index{antidot cluster} (on the left)
and of the reference sample (on the right).}
\label{fig:Afm_Adc}
\end{figure}
The micro-square with the 2$\times$2 antidot cluster\index{antidot cluster} consists of a
2$\times$2 $\mu m^2$
superconducting square with four antidots (i.e. square holes of 0.53
$\times$ 0.53 $\mu m^2$). A Pb/Cu bilayer with $50 \: nm$ of Pb and
$17 \: nm$
of Cu was used as the superconducting film for the fabrication of
this structure. The thin Cu layer was deposited on the Pb to 
protect it from oxidation and to provide a good contact-layer for 
the wire-bounding to the experimental apparatus. 
An AFM image of the Pb/Cu 2$\times$2 antidot cluster\index{antidot cluster}, is shown in
Fig.~\ref{fig:Afm_Adc} together with a reference sample\index{mesoscopic dot} (i.e. a Pb/Cu 
micro-square of  2$\times$2 $\mu m^2$  without antidots). 
The  Pb($50 \: nm$)/Cu($50 \: nm$) bilayer behaves as a Type II superconductor
with a $T_{c0}=6.05 \:K $, a coherence length\index{coherence length}, $\xi(0)\approx 35 \: nm$ and
a dirty limit penetration depth, $\lambda(0)\approx 76 \: nm$. The $T_c(H)$ 
measurements on the reference sample revealed characteristic features
originating from the confinement\index{confinement} of the superconducting condensate by 
the dot\index{mesoscopic dot} geometry (see Section II.C). The additional  features observed in 
the $T_c(H)$ phase boundary\index{superconducting phase boundary} of the antidot cluster\index{antidot cluster} can therefore be 
attributed to the presence of the antidots. 

\begin{figure}[h]
\centerline{\psfig{figure=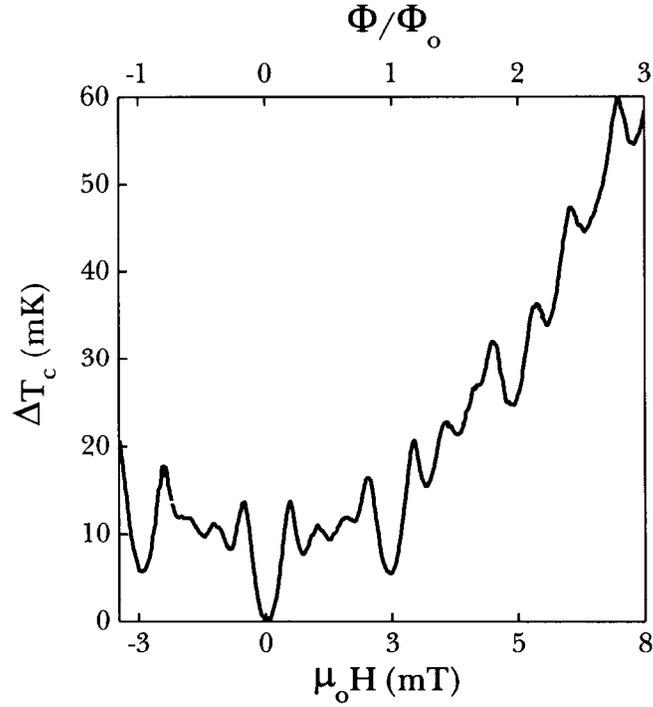}}
\caption{Experimental phase boundary, $\Delta T_c(H)$ for the
Pb/Cu 2$\times$2 antidot cluster\index{antidot cluster}.}
\label{fig:Psb1_Adc}
\end{figure}

The experimental $T_c(H)$ phase boundary\index{superconducting phase boundary} is shown in 
Fig.~\ref{fig:Psb1_Adc}. It was measured by keeping  the sample resistance
at 10\% of its normal state value and varying the magnetic field and
temperature\cite{Puig97}.
Strong oscillations are observed with a periodicity of 26 G and in each
of these periods, smaller dips appear at approximately 7.5 G, 13 G and
18 G. The parabolic background superimposed on $T_c(H)$
can again be described by Eq.~(\ref{TCBLine}).

Defining a flux quantum per
antidot as $\Phi_{0}=h/2e=BS$, where $B=\mu_{0}H$ and $S$ is an 
effective area per antidot cell ($S$= 0.8 $\mu m^2$), the minima observed
in the magnetoresistance and the  $T_c(H)$  phase boundary\index{superconducting phase boundary} at integer
multiples of 26 G can be correlated with a magnetic flux quantum per
antidot cell, $\Phi= n\Phi_{0}$. The ones observed at 7.5 G, 13 G 
and 18 G 
correspond to the values $\Phi/\Phi_{0}$=0.3, 0.5 and 0.7.

\begin{figure}[H]
\centerline{\psfig{figure=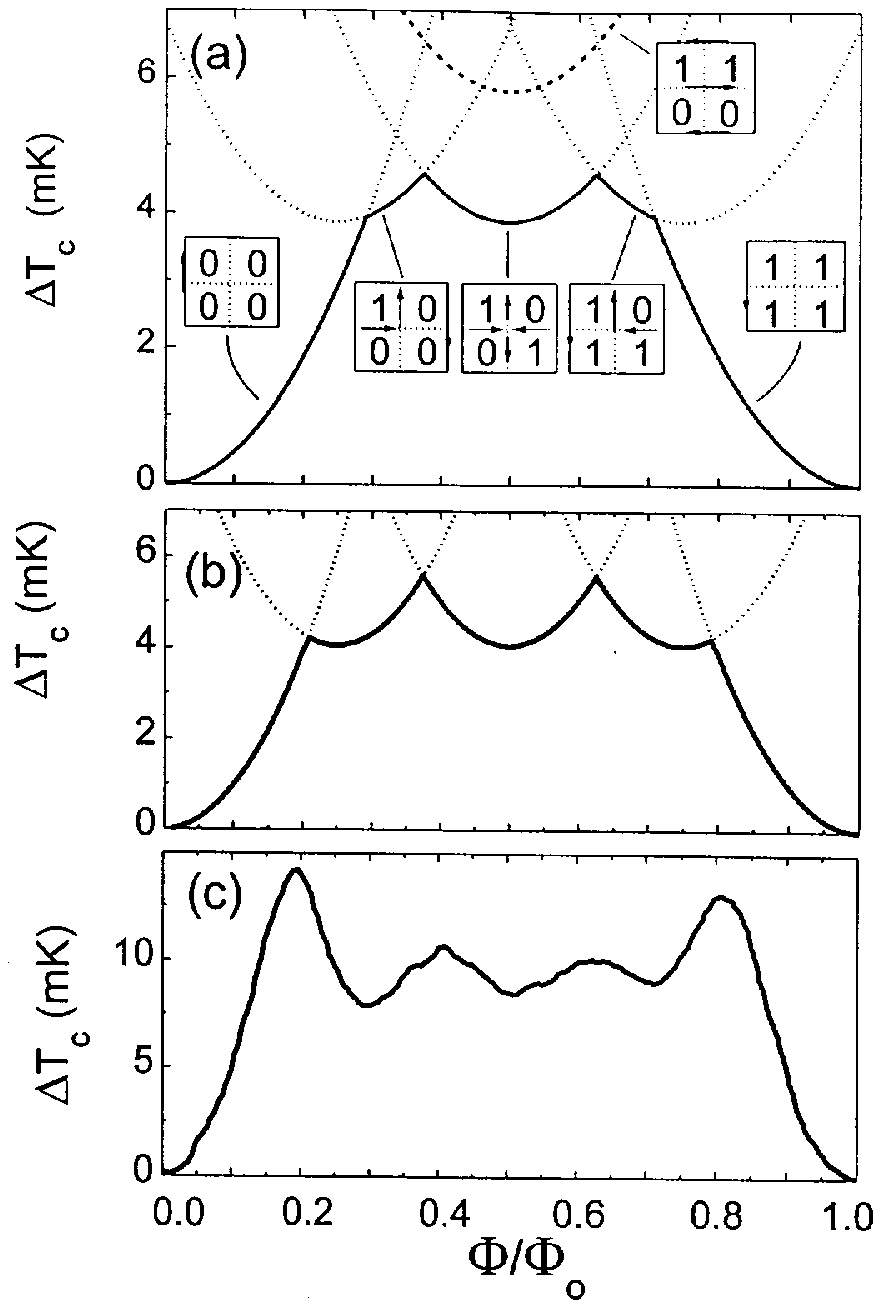}}
\caption{(a) Theoretical phase boundary, $T_c(\Phi/\Phi_{0})$,
calculated in the London limit of the 
Ginzburg-Landau theory without any fitting parameter (solid line). All
possible parabolic 
solutions are represented by dotted lines. The dashed line indicates the 
non-stable {\it parallel} 
configuration. The schematic representation of the $\{n_i\}$
quantum numbers at the antidots and 
characteristic current flow patterns for each parabolic branch are also
sketched. (b) The 
$T_c(\Phi/\Phi_{0})$ phase boundary, calculated as in
Fig.~\ref{fig:Psb2_Adc}a , but with the curvature "$\gamma$" of the
parabolae taken as a free parameter. The $\gamma$-value was increased by
a factor of two with 
respect to its calculated value used in Fig.~\ref{fig:Psb2_Adc}a.
(c) First period of the measured phase 
boundary shown in Fig.~\ref{fig:Psb1_Adc} after subtraction of the 
parabolic background. }

\label{fig:Psb2_Adc}
\end{figure}
The solutions obtained from the London model\index{London limit} (LL) define a phase boundary\index{superconducting phase boundary} which
is periodic in $\Phi$ with a periodicity of $\Phi_{0}$. Within each
parabola $\Delta T_c=\gamma(\Phi/\Phi_{0})^2$, where the coefficient
$\gamma$ characterizes the effective flux penetration trough the unit
cell. The $\gamma$-value is determined by the combination of $\lambda$
and the effective size of the current loops. 
In Fig.~\ref{fig:Psb2_Adc}, the first period of this phase boundary,
$\Delta T_c(\Phi)=T_{c0}-T_c(\Phi)$ versus $\Phi/\Phi_{0}$ is 
shown. Note that there are six parabolic solutions given by a different
set of flux quantum numbers $\{n_i\}$, each one defining a specific vortex
configuration\index{vortex configuration}. In Fig.~\ref{fig:Psb2_Adc}a, this is indicated by the 
numbers shown inside the schematic drawings of the antidot cluster\index{antidot cluster}. Note
that some vortex configurations\index{vortex configuration} are degenerate. 

From all these possible solutions, for each particular value of
$\Phi/\Phi_{0}$, only the branch with a minimum value of
$\Delta T_c(\Phi)$ is stable (indicated with a solid line in
Fig.~\ref{fig:Psb2_Adc}a). Note  that for the phase boundary\index{superconducting phase boundary},
calculated within the 1D model of 4 equivalent and properly attached
squares, no fitting parameters were used since the variation of
$T_c(\Phi)$ was calculated from the known values for $\xi$ and the size.
One period of the phase boundary\index{superconducting phase boundary} of the antidot cluster\index{antidot cluster} is composed
of five branches and in each branch a different stable vortex 
configuration\index{vortex configuration} is permitted. For the middle branch 
(0.37$<\Phi/\Phi_{0}<$0.63), the stable configuration is
the {\it diagonal} vortex configuration\index{vortex configuration} (antidots with equal
$n_i$ at the diagonals) instead of the {\it parallel} state 
(dashed line in Fig.~\ref{fig:Psb2_Adc}a).

The net supercurrent\index{supercurrent} density distribution circulating in the antidot
cluster for different 
values of $\Phi/\Phi_{0}$ has been determined using the same approach.
Circular currents flow around 
each antidot. For the states $n_i$=0 and $n_i$=1 currents flow in the opposite
direction, since currents 
corresponding to $n_i$=0 must screen the flux to fulfil the fluxoid 
quantization\index{fluxoid quantization} condition (Eq.~(\ref{Fluxoid})), 
whereas for $n_i$=1 they have to generate flux. At low values of
$\Phi/\Phi_{0}$, currents are canceled in 
the internal strips and screening currents only flow around the cluster.
When we enter the 
field range corresponding to the second branch of the phase boundary\index{superconducting phase boundary}, a 
vortex ($n_i$=1) is 
pinned around one antidot of the cluster (see Fig.~\ref{fig:Psb2_Adc}a).
At the third branch, the second 
vortex enters the structure and is localized in the diagonal. In the
fourth branch of the phase 
boundary the third vortex is pinned in the antidot cluster\index{antidot cluster}. And finally,
the current distribution 
for the fifth branch is similar to that of the first branch although 
currents flow in opposite 
direction.

Figure~\ref{fig:Psb2_Adc}c shows the first period of the measured phase
boundary $T_c(\Phi)$ after subtraction of 
the parabolic background. For all measured samples, the first period of 
the experimental 
phase boundary\index{superconducting phase boundary} is composed of five parabolic branches with minima at
$\Phi/\Phi_{0}$ = 0, 0.3, 0.5, 
0.7, 1. If we compare it with the theoretical prediction given in
Figure~\ref{fig:Psb2_Adc}a,  the overall shape 
can be reproduced although the experimental plot has two major peaks at 
$\Phi/\Phi_{0}$ = 0.2 and 0.8 whereas the theoretical curve only
predicts cusps around these positions.

The agreement between the measured and the calculated $T_c(\Phi)$ is improved if we assume that the coefficient $\gamma$ can be considered as
a fitting parameter. This 
seems to be feasible if we take into account the simplicity and limitation
of the used 1D 
model. Due to the relatively large width of the strips forming the 2$\times$2
cluster, the sizes of the current 
loops can change since they are "soft" in this case and not defined
very precisely.
 
As a result, the coefficient $\gamma$ can not be treated as a known
constant. If we use it as a 
free parameter (Fig.~\ref{fig:Psb2_Adc}b) then the curvature of all
parabolae forming $T_c(H)$ can be changed 
and the calculated $T_c(H)$ curve becomes closer to the experimental one
though the amplitude 
of the maxima at $\Phi/\Phi_{0}$ = 0.2 and 0.8 is still lower than 
in the experiment (Fig.~\ref{fig:Psb2_Adc}c). 
The discrepancy in the description of the amplitude of the maxima at
$\Phi/\Phi_{0}$= 0.2 and 0.8 could also be related to the pinning\index{vortex pinning} 
of vortices by the antidot cluster\index{antidot cluster} when potential barriers between 
different vortex configurations\index{vortex configuration} may appear. At the same time, the 
achieved agreement between the positions of the measured and calculated
minima of the $T_c(H)$ curves confirms that the observed effects are
due to fluxoid quantization\index{fluxoid quantization} and formation of certain stable vortex 
configurations\index{vortex configuration} at the antidots.

\section{Superconducting films with an antidot lattice\index{antidot lattice}}
Laterally nanostructured superconducting films having regular arrays of
antidots\index{antidot array} are convenient model objects to study the effects of the
confinement\index{confinement} topology on the $T_c(H)$ phase boundary in two different
regimes. The first (or "collective") regime corresponds to the 
situation where all elements A, forming an array, are coupled. 

From the experimental $T_c(H)$ data on antidot clusters\index{antidot cluster} we expect for
films with an antidot lattice\index{antidot lattice}, higher critical fields at
$\Phi=n\Phi_{0}$, which is in agreement with the appearance of the
$T_c(H)$ cusps at  $\Phi=n\Phi_{0}$ in superconducting
networks\index{superconducting wire networks} \cite{Pannetier91}. Here, the flux $\Phi$ is calculated per unit
cell of the antidot lattice\index{antidot lattice}.

On the other hand, by applying sufficiently high magnetic fields, the
individual circular currents flowing around antidots, can be decoupled
and the crossover to a "single object" behaviour could be observed. In 
this case the relevant area for the flux is the area of the antidot 
itself and we deal with surface superconductivity\index{surface superconductivity} around an antidot. 

\begin{figure}[H]
\centerline{\psfig{figure=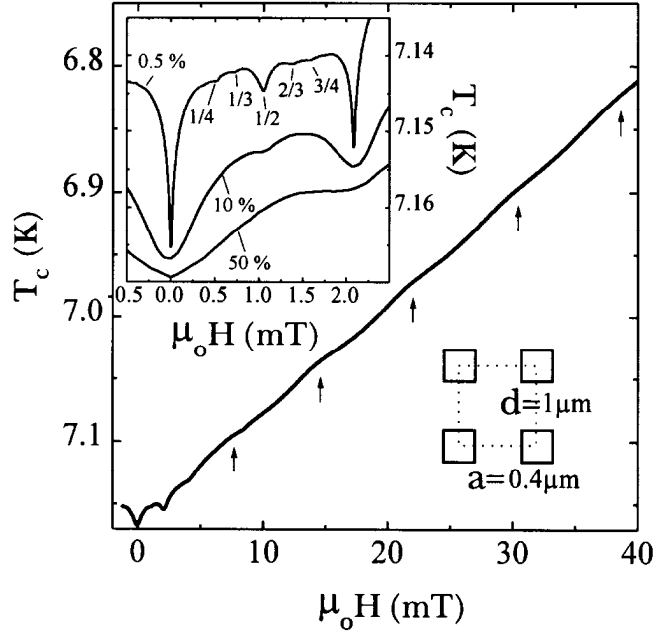}}
\caption{Critical field of a Pb($50 \: nm$) film measured at 10\%~$R_n$, 
with $d=1\: \mu m,\: a=0.4 \: \mu m$. The inset shows a zoom of the first
cusp at different criteria 50\%~$R_n$, 10\%~$R_n$ and 0.5\%~$R_n$.}

\label{fig:Psb_Pb}
\end{figure}

Figure~\ref{fig:Psb_Pb} shows the critical field for a Pb($50 \: nm$) sample
with a square antidot lattice\index{antidot lattice} (period  $d=1 \: \mu m$ and the antidot size
$a=0.4 \: \mu m$). The $T_c(H)$ boundary is determined at 10\% of the
normal state resistance,  $R_n$. In this graph two distinct 
periodicities are present. 

\begin{figure}[h]
\centerline{\psfig{figure=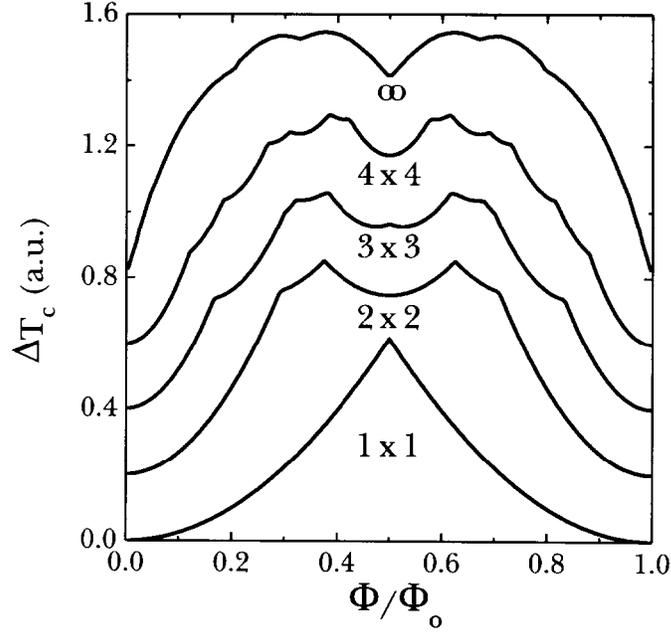}}
\caption{Calculations of the first $T_c(H)$ period for an
$N\times\,N$ antidot system ($N=1,2,3,4,\infty$) in the London limit.}

\label{fig:Psb_NxN}
\end{figure}

Below $\sim$ 8~mT cusps are found with a period of 2.07 mT, corresponding 
to one flux-quantum per lattice cell. These cusps or "collective
oscillations" \cite{Bezryadin96a} are reminiscent of superconducting wire
networks\index{superconducting wire networks} \cite{Pannetier91} and arise from the phase correlations between
the different loops which constitute the network.
\noindent
These cusps are obtained by narrowing the minima at $n\Phi_{0}$ with
increasing size $N$ of the $N\times N$ antidot cluster\index{antidot cluster} (see the
sharpening of the minima at $\Phi=0,\,\Phi_{0}$ in 
Fig.~\ref{fig:Psb_NxN};
note that the phase boundary\index{superconducting phase boundary} in the $N\rightarrow \infty$ case
has a similar shape as the lowest energy level of the
Hofstadter butterfly\index{Hofstadter butterfly} \cite{Pannetier84,Hofstadter76}). 
\noindent
An important observation  is that the amplitude of these "collective"
oscillations depends upon the choice of the resistive criterion. This
is similar to the case of  Josephson junction arrays and weakly coupled
wire networks\index{superconducting wire networks} \cite{Giroud92} where phase fluctuations  dominate the
resistive behavior. The inset of Fig.~\ref{fig:Psb_Pb} shows the
first collective period, measured using  three different criteria.
As the criterion is lowered  the cusps become sharper and the
amplitude increases well above the prediction based on the mean 
field theory for strongly coupled wire networks\index{superconducting wire networks} \cite{Pannetier91}.
At the same time, cusps appear at rational fields 
$\Phi/\Phi_{0}$=1/4, 1/3, 1/2, 2/3 and 3/4 arising  from the
commensurability\index{commensurability} of the vortex structure with the underlying 
lattice. 

Above $\sim$ 8~mT, the collective oscillations die out and "single object"
cusps appear, having a periodicity which roughly corresponds to one flux
quantum $\Phi_{0}$ per antidot area, $a^2$.
These cusps are due to the transition between localized superconducting
edge states \cite{Bezryadin96a}  having a different angular momentum\index{orbital quantum number} $L$.
These states are formed around the antidots and are described by the
same orbital momentum introduced in Section~II.C for dots.    
\begin{figure}[H]
\centerline{\psfig{figure=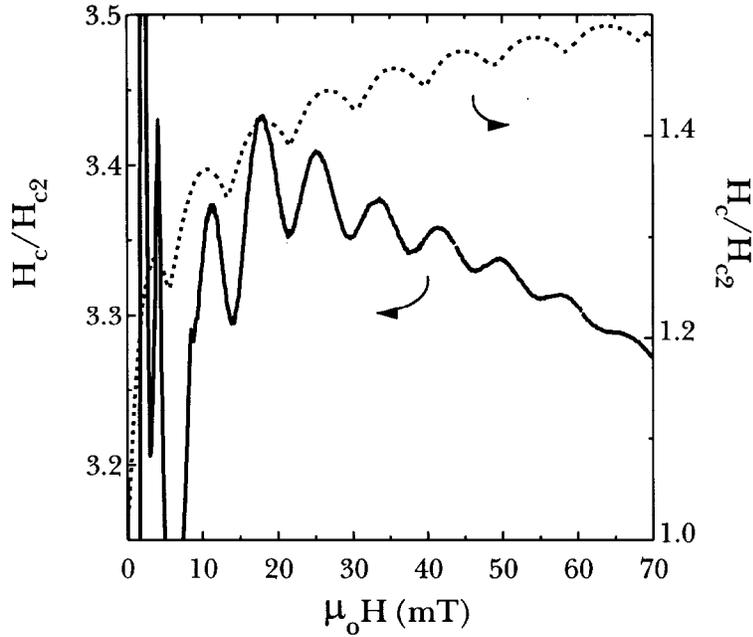}}
\caption{The critical field of Fig.~\ref{fig:Psb_Pb}  normalized by
$\Phi_{0}/2\pi\xi^2(T)$ versus the applied field. The dashed line 
(right axis) shows the theoretical result for a single circular hole
with a radius $R=0.24\, \mu m$.}

\label{fig:Red_Psb_Pb}
\end{figure}

Figure~\ref{fig:Red_Psb_Pb} shows the same critical field as shown in
Fig.~\ref{fig:Psb_Pb}, but normalized
by the upper critical field $H_{c2}$ of a plain film without antidots, 
$\mu_{0}H_{c2}=\Phi_{0}/2\pi\xi^2(T)$ ( $\xi(0)=36 \: \: nm$).
The dashed line is the calculation of  the reduced critical field 
for a plain film with a single circular antidot with radius $R=0.24 \: \mu m$.
The positions of the cusps correspond reasonably well to the experimental
ones, taking into account that the model only considers a single hole. From
this comparison, an effective area $\pi R^2 =0.187\:\mu m^2$ is determined
which is close to the experimental value $a^2=0.16\:\mu m^2$.  

From Figures~\ref{fig:Psb_Pb} and~\ref{fig:Red_Psb_Pb} it is possible to
show that the transition from the network regime to the "single object"
regime takes place at a temperature $T^*$  given by the relation 
$w \approx 1.6\,\xi(T^*)$, (where $w$ is the width of the superconducting 
region between two adjacent antidots). 

Experiments on systems with other antidot sizes demonstrate that the $a/d$
ratio determines the relative importance of the "collective regime" and
changes the cross-over temperature $T^*$. The relation 
$w \approx 1.6\:\xi(T^*)$, seems nevertheless to hold reasonably well
and is similar to the transition from bulk nucleation of
superconductivity to surface nucleation\index{surface superconductivity} in a thin superconducting
slab parallel to the magnetic field \cite{Saint-James65}, which happens
at a temperature $T_{cr}$ satisfying $w=1.8 \: \xi(T_{cr})$.

Comparing the bulk $H_{c2}(T)$ curve with the $T_c(H)$ boundary for films 
with an antidot lattice\index{antidot lattice}, we clearly see a qualitative difference between the
two, caused by the lateral micro-structuring. In the network limit,
$T_c(H)$ can be related to the lowest $E_{LLL}$ level in the Hofstadter
butterfly\index{Hofstadter butterfly} \cite{Pannetier84,Hofstadter76} with pronounced cusps at
$n\Phi_{0}$ and the substructure within each period. Reducing the
size of the antidot, we are modifying $T_c(H)$ substantially but the 
cusps at $n\Phi_{0}$ are still clearly seen \cite{Rosseel97}. 

Finally, by introducing antidot lattices\index{antidot lattice} we stabilize the novel flux
states, which can be briefly characterized as follows. For relatively large
antidots sharp cusp-like magnetization anomalies appear at the matching
fields $H_m$. These anomalies are caused by the formation of the
multi-quanta vortex lattices\index{multi-quanta vortex lattice} at each subsequent
$H_m$\cite{Baert95a,Baert95b,Moshchalkov96,Bezryadin96b}. The multi-quanta
vortex lattices\index{multi-quanta vortex lattice} make possible a peaceful coexistence of the flux
penetration at the antidots and the presence of a substantial superfluid
density in the space between them. This leads to a very strong enhancement
of the critical current density in films with an antidot lattice\index{antidot lattice}. For
smaller antidots the vortices are forced to occupy the interstitial
positions after the saturation of the pinning\index{vortex pinning} sites at
antidots\cite{Bezryadin96b,Rosseel96,Harada96}. This leads to the
formation of the novel composite flux-line lattices consisting from
the interpenetrating sublattices of weakly pinned interstitial
single-quantum vortices and multi-quanta vortices\index{multi-quanta vortex lattice} strongly pinned by the
antidots. When the interstitial flux-line lattice melts, it forms the
interstitial flux-liquid coexisting with the flux solid at the antidots. 

\section{Conclusions}
We have carried out a systematic study of \index{confinement} and quantization\index{quantization}
phenomena in nanostructured superconductors.  The main idea of
this study was to vary the boundary conditions\index{boundary conditions} for confining the
superconducting condensate by taking samples of different
topology and, through that, to modify the lowest Landau level\index{Landau levels} $E_{LLL}(H)$
and therefore the critical temperature $T_{c}(H)$.  Three different types
of samples were used: (i) individual nanostructures\index{nanostructures} (lines, loops, dots),
(ii) clusters of nanoscopic elements - 1D clusters of loops and 2D
clusters of antidots, and (iii) films with huge regular arrays of antidots
(antidot lattices\index{antidot lattice}). We have shown that in all these structures, the phase
boundary\index{superconducting phase boundary} $T_{c}(H)$ changes dramatically when the confinement topology for
the superconducting condensate is varied.  The induced $T_{c}(H)$ variation
is very well decribed by the calculations of $E_{LLL}(H)$ taking into
account the imposed boundary conditions\index{boundary conditions}.  These results convincingly
demonstrate that the phase boundary\index{superconducting phase boundary}in $T_{c}(H)$ of nanostructured
superconductors differs drastically from that of corresponding bulk
materials. Moreover, since, for a known geometry $E_{LLL}(H)$ can
be calculated a priori, the superconducting critical parameters,
i.e. $T_{c}(H)$, can be controlled by designing a proper confinement\index{confinement}
geometry. While the optimization of the superconducting critical
parameters has been done mostly by looking for different
materials, we now have a unique alternative - to improve the superconducting
critical parameters of {\it the same material} through the optimization of
{\it the confinement topology} for the superconducting condensate and for
the penetrating magnetic flux. 

\section*{Acknowledgments}

The authors would like to thank A. L\'{o}pez and V.~Fomin for fruitful
discussions and R.~Jonckheere for the electron beam lithography.
We are grateful to the Flemish Fund for Scientific Research (FWO),
the Flemish Concerted Action (GOA), the Belgian Inter-University 
Attraction Poles (IUAP) and the European Human Capital and Mobility (HCM)
research programs for the financial support.  E. Rosseel is a Research 
Fellow of the Belgian Interuniversity Institute for Nuclear Sciences
(I.I.K.W.), and L. Van Look of the European project JOVIAL.  
M. Baert is a Postdoctoral 
Fellow supported by the Research Council of the K.U.Leuven.

\pagebreak

\end{document}